\documentclass{article}

\usepackage[final]{ml4mols_2020}

\usepackage[utf8]{inputenc} 
\usepackage[T1]{fontenc}    
\usepackage{hyperref}       
\usepackage{url}            
\usepackage{booktabs}       
\usepackage{amsfonts}       
\usepackage{nicefrac}       
\usepackage{microtype} 

\usepackage{sidecap}
\usepackage{amssymb}
\usepackage[ruled,vlined]{algorithm2e}
\usepackage{longtable}
\usepackage{amsmath}
\usepackage{hyperref}
\usepackage{array}
\usepackage{subfig}
\usepackage{rotating}
\usepackage{pgfgantt}
\usepackage[T1]{fontenc}
\usepackage[utf8]{inputenc}
\usepackage{wrapfig}
\usepackage[autostyle]{csquotes}

\title{Design of Experiments \\ for Verifying Biomolecular Networks}

\author{Ruby Sedgwick\textsuperscript{1,2} \hspace{0.22cm} John Goertz\textsuperscript{1} \hspace{0.21cm}  Molly Stevens\textsuperscript{1} \hspace{0.21cm}  Ruth Misener\textsuperscript{2}  \hspace{0.21cm} Mark van der Wilk\textsuperscript{2} \\
 \textsuperscript{1}Department of Materials, Department of Bioengineering  and Institute of Biomedical Engineering \\
  \textsuperscript{2}Department of Computing \\
  Imperial College London \\
  \texttt{\{r.sedgwick19, j.goertz, m.stevens, r.misener, m.vdwilk\}@imperial.ac.uk}
}

\begin{document}

\maketitle

\begin{abstract}
There is a growing trend in molecular and synthetic biology of using mechanistic (non machine learning) models to design biomolecular networks. Once designed, these networks need to be validated by experimental results to ensure the theoretical network correctly models the true system. However, these experiments can be expensive and time consuming. We propose a design of experiments approach for validating these networks efficiently. Gaussian processes are used to construct a probabilistic model of the discrepancy between experimental results and the designed response, then a Bayesian optimization strategy used to select the next sample points. We compare different design criteria and develop a stopping criterion based on a metric that quantifies this discrepancy over the whole surface, and its uncertainty. We test our strategy on simulated data from computer models of biochemical processes. 
\end{abstract}

\section{Introduction}
As mathematical models of biological systems improve, they are increasingly used to design systems for a particular purpose.
Our motivating example is the creation of bespoke networks of interacting biomolecules that respond to changes of input concentrations as logic gates (e.g.~AND or OR). These can be used to create biosensors  \cite{siuti_synthetic_2013}, targeted drug delivery \cite{badeau_engineered_2018}, and tunable genetic circuits \cite{bartoli_tunable_2020}.
Due to discrepancies between the predictions of the mathematical model and biological reality, each network has to be validated experimentally before use to verify that the real-world system exhibits the desired response.
In the context of verifying many networks, the cost of these experiments becomes large. In this work, we propose a method for verifying whether the true response of the network is suitably similar to that which is desired. 

The response of biomolecular networks have been validated before \cite{siuti_synthetic_2013, bashor_complex_2019, schaerli_unified_2014}, by running experiments at a fixed predetermined set of input points that were deemed important for the application. The number of points was chosen to satisfy the experimenter that the surfaces were similar. A similar, but significantly different problem, is investigated in the surrogate modeling literature, where it is also important to understand the discrepancy between experimental data and a true reference \cite{atamturktur_defining_2015, williams_batch_2011, unal_improved_2011, moore_batch_2009}. We take inspiration from these methods in (1) modeling the discrepancy so the observed data informs our prediction of the discrepancy over the entire surface and (2) using the discrepancy model to intelligently determine the next experiment to run. In our work we follow a similar procedure; each time more experimental data is obtained, a Gaussian process is fitted to the discrepancy between the experimental data and the computer model. We use standard Bayesian optimization design criteria to select the next experimental point, although this element of our method could be improved. The Gaussian process is then updated, before the process repeats. Our main contribution is in the proposal of a principled stopping criterion designed to recognize when enough experiments have been conducted to accept or reject a network. We assessed the performance of this method by applying it to the validation of simulated biomolecular networks, although this method could be useful for any application where it is necessary to test whether a designed response surface matches experimental reality.

\section{Methods}

We wish to reject an incorrect (or accept a correct) biomolecular network as quickly as possible. To do this, our method can be broken down into: (i) modeling the discrepancy between the theoretical assay design and the true underlying response surface, (ii) the stopping criterion and (iii) the design criterion.

\subsection{Discrepancy Modeling}

To understand how well the experimental data fits the designed surface, we model the discrepancy \(g(\boldsymbol{x})\), \(g : \mathbb{R}^D \mapsto \mathbb{R}\) as in Eq.~\ref{eq:observation breakdown} \cite{kennedy_bayesian_2001}. In an input domain \(\mathcal{X} \in\mathbb{R}^D\), where \(D\) is the dimensions of the input domain, for an input  \(\boldsymbol{x}\in\mathcal{X}\) the experimental observations \(y(\boldsymbol{x})\), \(y : \mathbb{R}^D \mapsto \mathbb{R}\), can be split into the model prediction \(\eta(\boldsymbol{x}, \boldsymbol{\theta_m})\), \(\eta : \mathbb{R}^D \mapsto \mathbb{R}\), with best-fit parameter values, where \(\boldsymbol{\theta}_m\) is a vector of the mechanistic model parameters; the mechanistic model inadequacy \(\psi(\boldsymbol{x})\), \(\psi : \mathbb{R}^D \mapsto \mathbb{R}\), which is unknown in untested conditions; and the experimental noise \(\epsilon(\boldsymbol{x})\), \(\epsilon : \mathbb{R}^D \mapsto \mathbb{R}\). This breakdown allows for explicit modeling of the inadequacy term \(\psi(\boldsymbol{x})\) to account for the systematic deviations from the model. 

\begin{equation}
    g(\boldsymbol{x}) =  y(\boldsymbol{x}) - \eta(\boldsymbol{x}, \boldsymbol{\theta}_m) = \psi(\boldsymbol{x}) + \epsilon(\boldsymbol{x})
     \label{eq:observation breakdown}
\end{equation}

 We use a squared exponential Gaussian process as a probabilistic model for the discrepancy, implemented using GPflow \cite{de_g_matthews_gpflow_2017}. To better understand how the algorithm is working, we simulate the experimental data as noiseless, so therefore set the noise variance to be very small \(\sigma_n^2 = 1e^{-4}\) and untrainable. We set both the signal variance and lengthscale priors to a gamma distribution with a high mode \(\sigma_f^2 \sim Gamma(2, 1)\). To train the model, the most probable values of the hyperparameters are found using the maximum a posteriori (MAP) estimate. A maximum variance design criterion was used to select the next sample point. 

\subsection{Stopping Criterion} \label{section:Stopping Criterion}

The aim of the stopping criterion is to terminate the process when a reasonable degree of certainty that the network architecture should be accepted or rejected has been reached. To do this, we need a measure of the average discrepancy over the entire surface. Only the absolute discrepancy matters, regardless of its sign and a measure of certainty in the estimates is required. Therefore, we developed a stopping metric based on the RMSE of the surface.

As the Gaussian process is fitted to the true discrepancy, the stopping metric needs to convert the true discrepancy into an absolute value. The metric should also incorporate the uncertainty in the model, to prevent a decision from being made on the basis of too little information. We developed a root mean squared error (RMSE) based metric \(M\) to quantify the overall discrepancy. \(N\) function samples \(\boldsymbol{f}_{s} = \{f_{s}(\boldsymbol{x_{j*}})\}_{j=0}^m\) are drawn from the Gaussian process posterior \(f_*\) for \(m\) sample inputs \(\boldsymbol{X}_* = \{\boldsymbol{x}_{j*}\}_{j=0}^{m}\) spaced uniformly across the input domain, such that \(\boldsymbol{F}_s = \{\boldsymbol{f}_{sh}\}_{h=0}^{N}\) . The RMSE metric for each Gaussian process sample is calculated as in Eq.~\ref{eq: Metric_def}.
\begin{equation}
    M_h = \sqrt{\frac{1}{m}\sum_{j*=0}^{m} f_{sh}(\boldsymbol{x}_{j*})^2}
    \label{eq: Metric_def}
\end{equation}

\begin{figure}[hbt]
 
    \begin{minipage}[b]{0.41\textwidth}
    \includegraphics[scale=0.38]{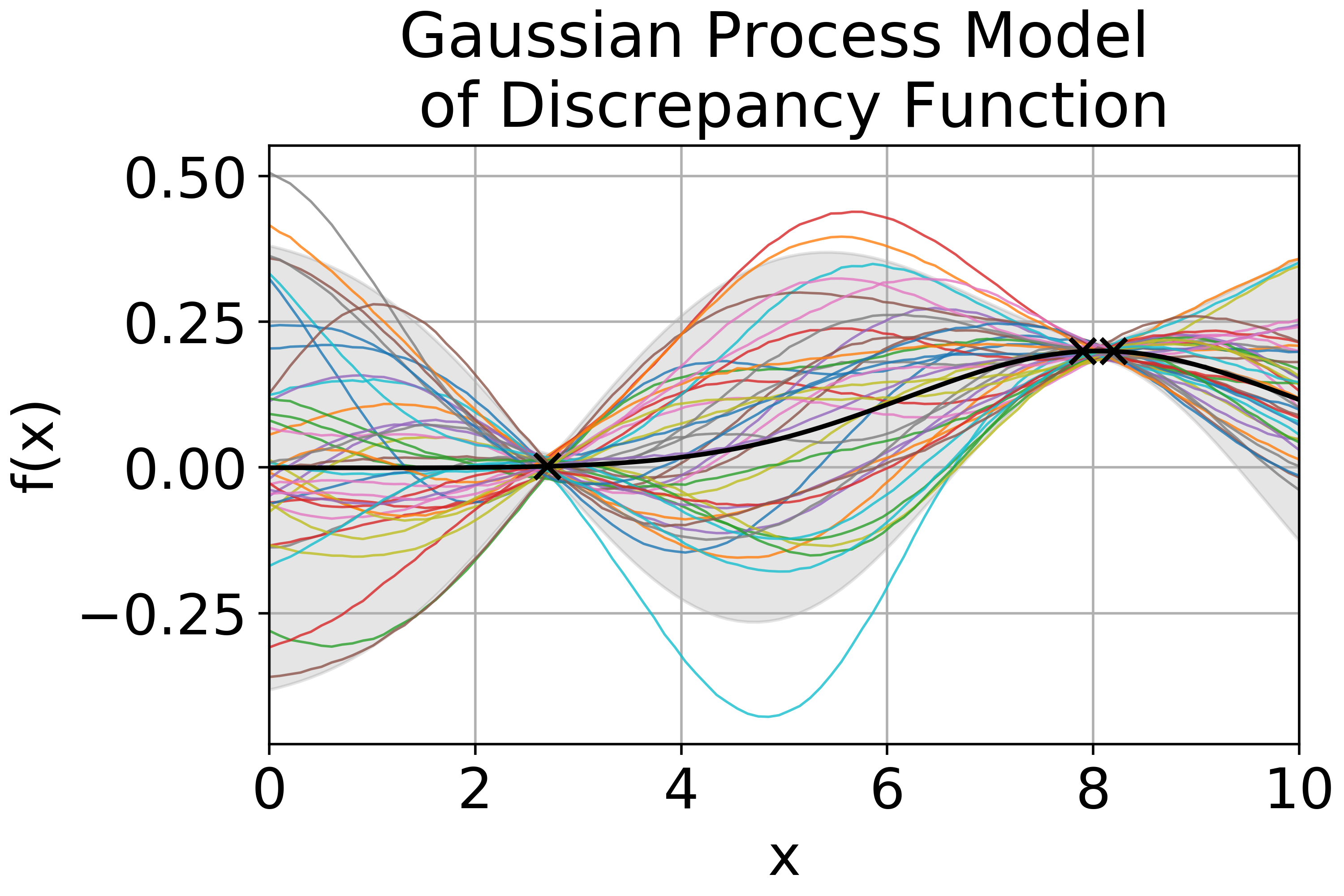}
    \end{minipage}
    \begin{minipage}[b]{0.29\textwidth}
    \includegraphics[scale=0.38]{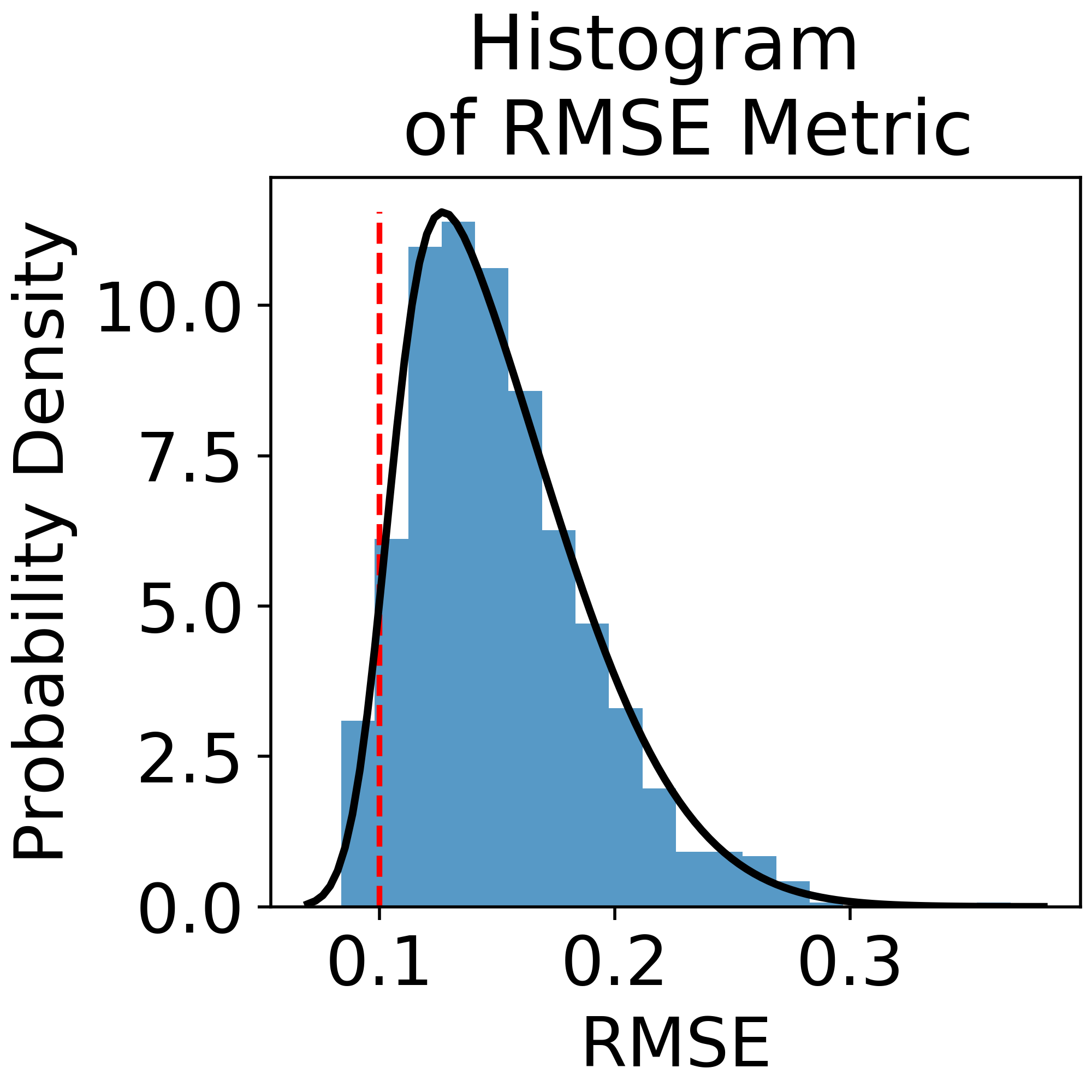}
    \end{minipage}
    \begin{minipage}[b]{0.29\textwidth}
    \caption{\footnotesize Plot of samples drawn from a one dimensional Gaussian process posterior (left) and a histogram of the distribution of metrics \(M\) for 40 samples drawn from this Gaussian process (right), with a skewed normal distribution fitted to the data (black line). The RMSE threshold \(T_{avg}\) is marked with the dashed red line.}
    \label{fig:metric_dist}
    \end{minipage}
    
\end{figure}

This metric directly measures the average discrepancy of the surface and allows for the propagation of uncertainty from the probabilistic model of the surface to the metric itself. Calculating this metric for numerous function draws gives a distribution of metric values as shown in Fig.~\ref{fig:metric_dist}. The probability \(p(M_{true} < T_{avg})\) that the true RMSE of the deviation lies below the threshold for average deviation \(T_{avg}\) set by a domain expert is then calculated. If the probability is smaller than a predetermined minimum probability \(p_{reject}\) or larger than a predetermined maximum probability \(p_{accept}\), then the network architecture is rejected or accepted respectively. 
If the network architecture is not accepted or rejected, then the process continues. MAP inference for the GP hyperparameters has a tendency to cause predictions to be over-confident for very small sets of observations. To prevent premature termination, the stopping criterion is only applied after a given number of data points have been observed.

\subsection{Design Criterion}

Four Bayesian optimization design criteria; maximum variance, expected improvement and upper confidence bound with weightings of 1 and 3 on the standard deviation, were compared to three standard experimental strategies: grid search, random and Sobol random. Each strategy was tested multiple times on 8 test surfaces, and the absolute error in mean and maximum discrepancy recorded (Figure~\ref{fig:design_crit_test_all_avg} in Appendix~\ref{ap:DesignCriteria}). None of the regimes tested outperformed the others for both maximum and average discrepancies, but maximum variance, UCB with high weighting on variance and Sobol methods were found to perform the best. 

\section{Results}

We ran tests to determine how accurately, and in how many runs, the stopping metric can determine if a designed network should be accepted or rejected. 12 simulated biochemical networks were compared to the ideal logic gate surfaces they were designed to recreate. We ran the algorithm 10 times for each network to account for the semi-random nature of the acquisition function. Figure~\ref{fig:sequential} in Appendix~\ref{ap:biochem} demonstrates the sequential procedure of the method. The ideal AND, OR and XOR surfaces provide a +1/-1 response based on the input concentrations of biomolecules. The RMSE metrics at each iteration were recorded so that the stopping metric could be applied retrospectively, meaning all tests were done on the same dataset. The two stopping metric parameters that we explored are the minimum number of experiments before the stopping metric is applied and the percentage certainty we wish to have before accepting or rejecting a network.

\begin{figure}
 \centering
    \includegraphics[scale=0.55]{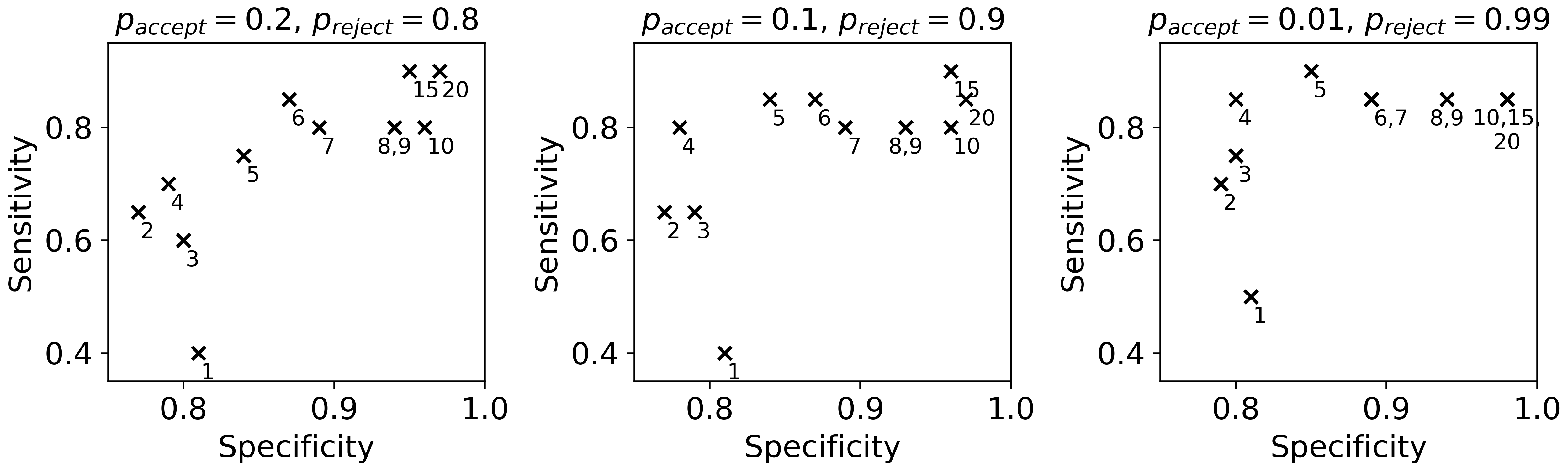}
    \caption{\footnotesize Sensitivity-specificity plots for different levels of certainty \(p_{accept}\) and \(p_{reject}\). The numbers next to the points are the minimum number of runs.}
    \label{fig:sens_spec}
\end{figure}

To investigate the effect of the minimum number of experiments, we measured the numbers of false acceptances/rejections for a minimum number of experiments \(n_{min} = \{1,2,3,..,10, 15, 20\}\). The RMSE threshold \(T_{avg}\) was set by a domain expert to \(0.2\), as this was deemed an acceptable level of discrepancy.  Figure~\ref{fig:sens_spec} shows a plot of sensitivity against specificity for different values of \(n_{min}\) and different percentage certainties. These results show that there is a trade off between the minimum number of experiments and the number of false acceptances and rejections made by the algorithm. This is because for larger minimum number of experiments the algorithm has seen more data points so has more information on what the discrepancy surface looks like.

\begin{figure}
    \centering
    \begin{minipage}[b]{0.32\textwidth}
     \includegraphics[scale=0.38]{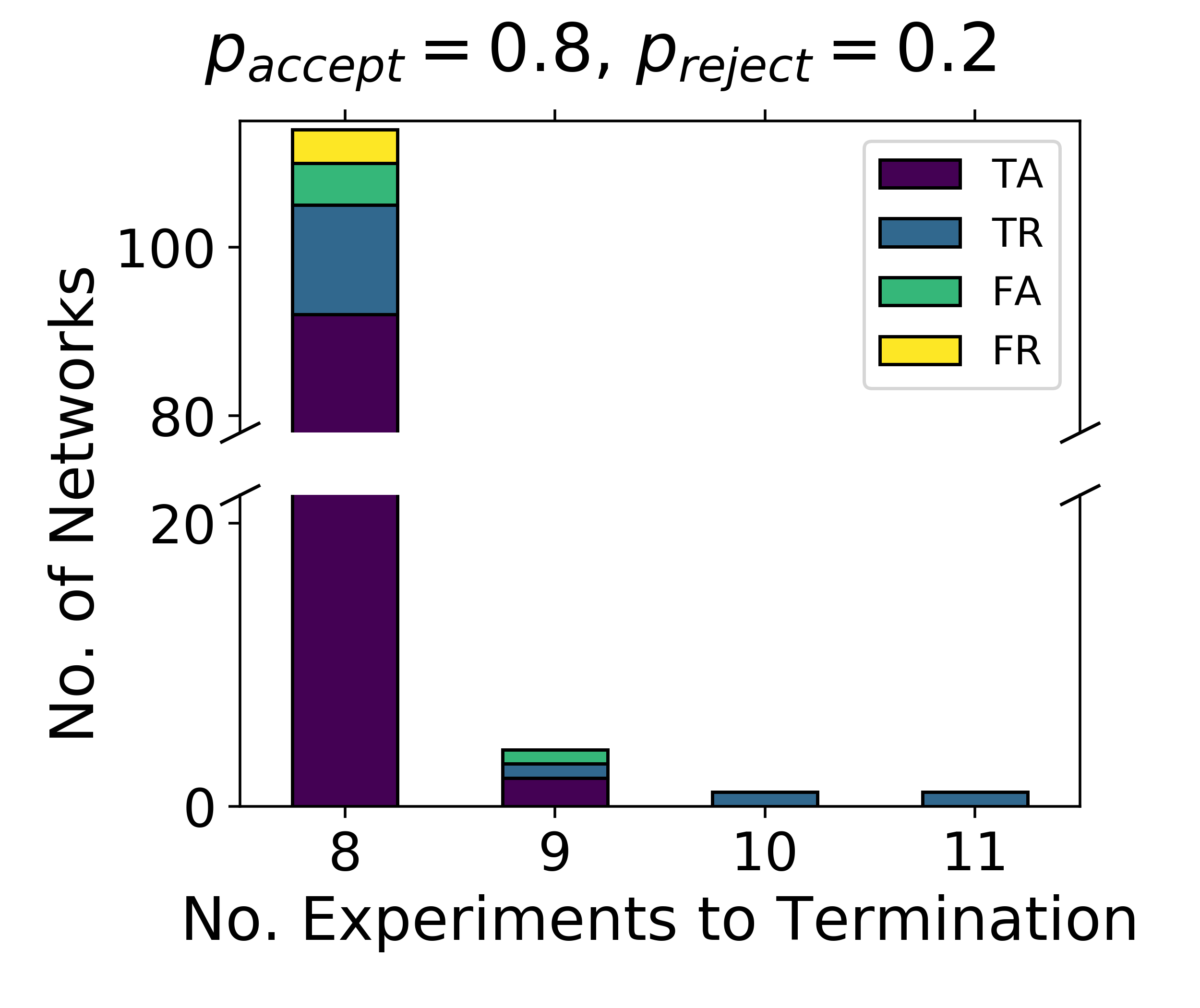}
    \end{minipage}
    \begin{minipage}[b]{0.32\textwidth}
    \includegraphics[scale=0.38]{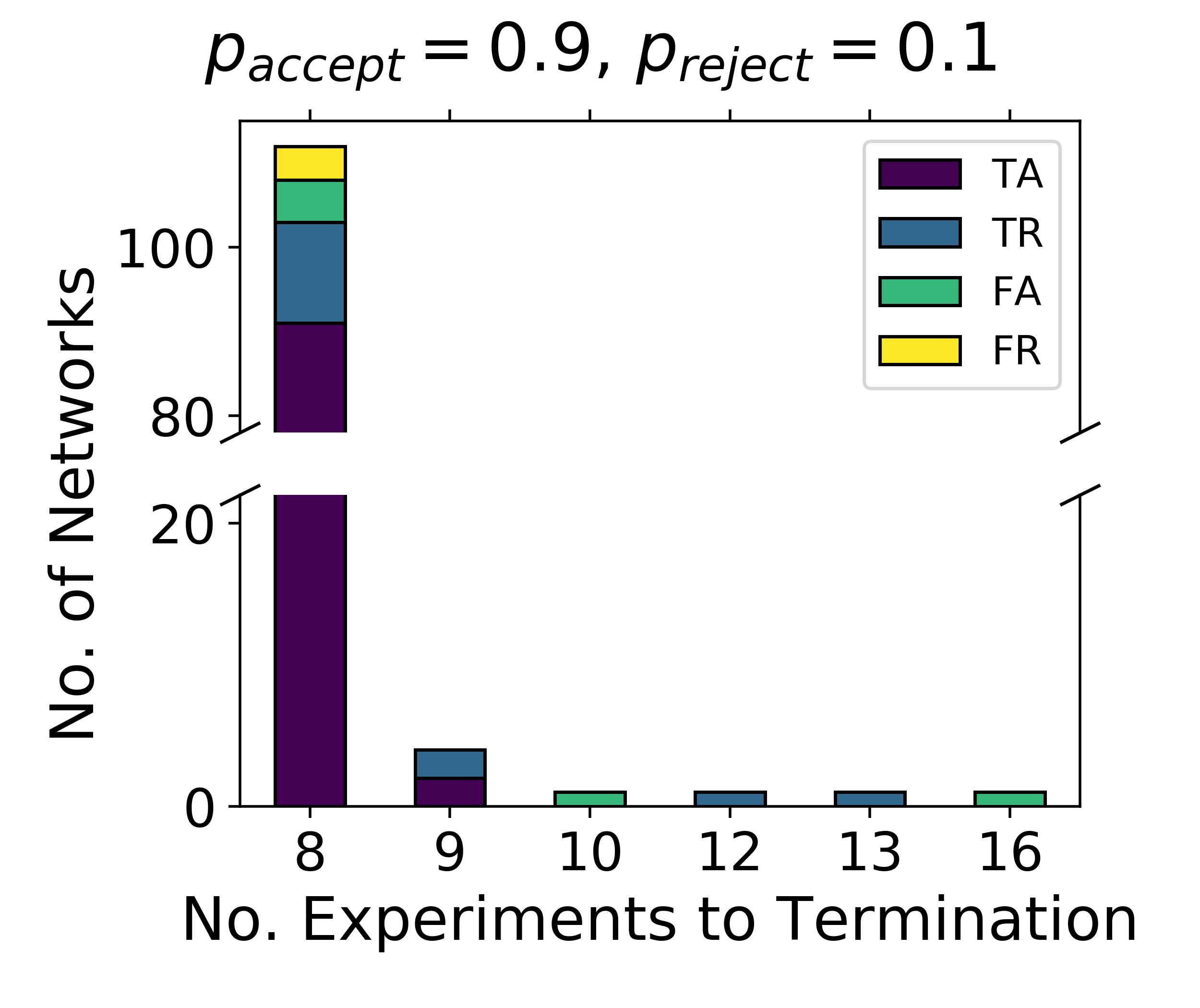}
    \end{minipage}
    \begin{minipage}[b]{0.32\textwidth}
    \includegraphics[scale=0.38]{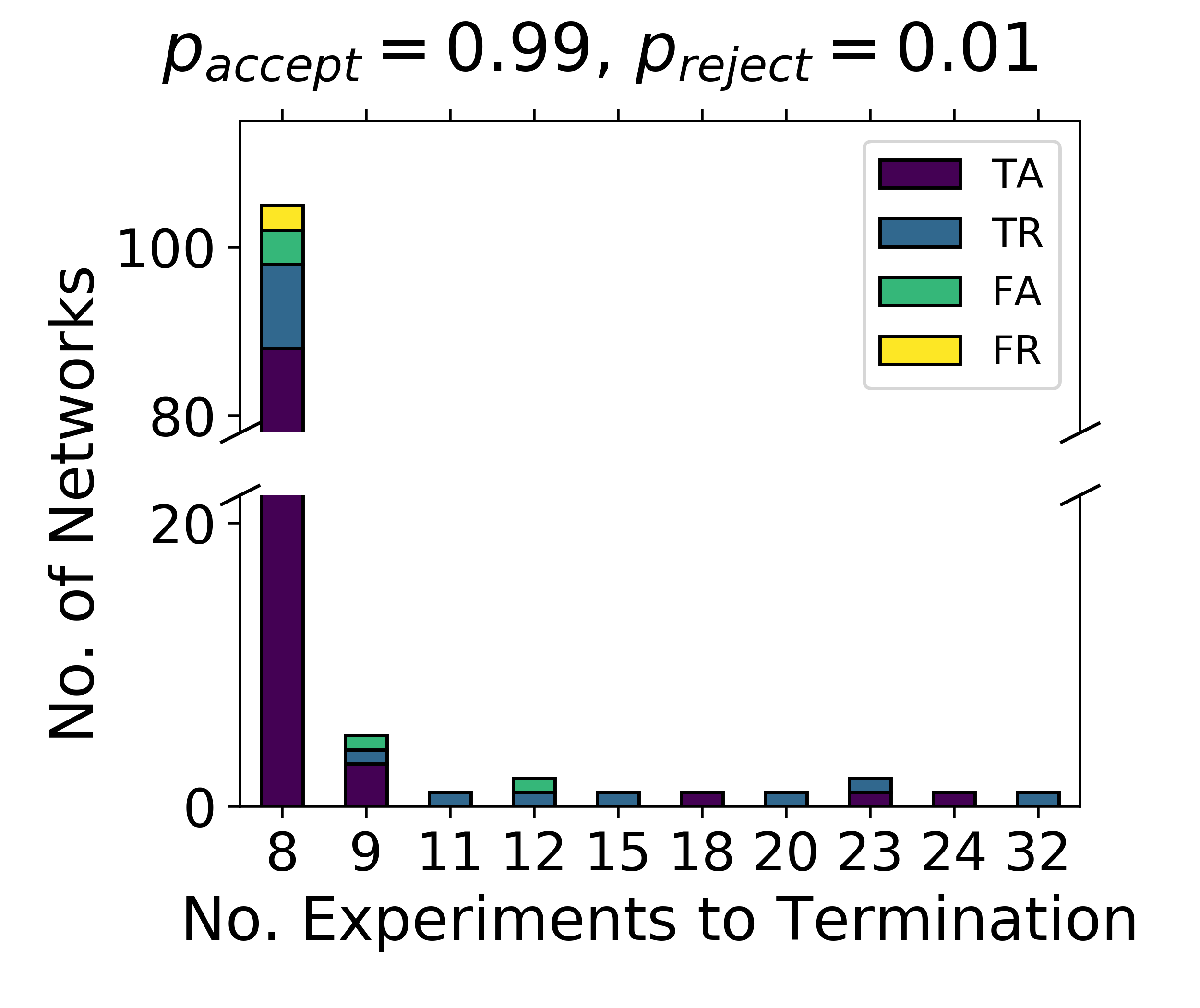}
    \end{minipage}
    \vspace{-10pt}

    \caption{\footnotesize The number of experiments to termination of the experimental run for different levels of confidence in the result. These results are split by outcomes: true acceptance (TA), true rejection (TR), false acceptance (FA) and false rejection (FR). }
    \label{fig:experiments_term}
\end{figure}

We chose the optimal minimum number of experiments to be \(n_{min} = 8\) as the improvement in sensitivity and specificity is small for values of \(n_{min}\) greater than this. Figure~\ref{fig:experiments_term} shows when terminations of the algorithm occur and the type of termination for  \(n_{min} = 8\) for 3 levels of certainty we wish to have in our results, \(p_{accept}=\{0.8, 0.9, 0.99\}\) and \(p_{reject}=\{0.2, 0.1, 0.01\}\). Increasing the level of certainty required before accepting or rejecting a network increases the number of experiments before termination. When choosing the value for \(p_{accept}\) and \(p_{reject}\), the tolerance for erroneously accepting or rejecting should take the application into consideration.

\begin{wrapfigure}{r}{0.3\textwidth}
    \begin{center}
    \vspace{-50pt}
    \includegraphics[scale=0.4]{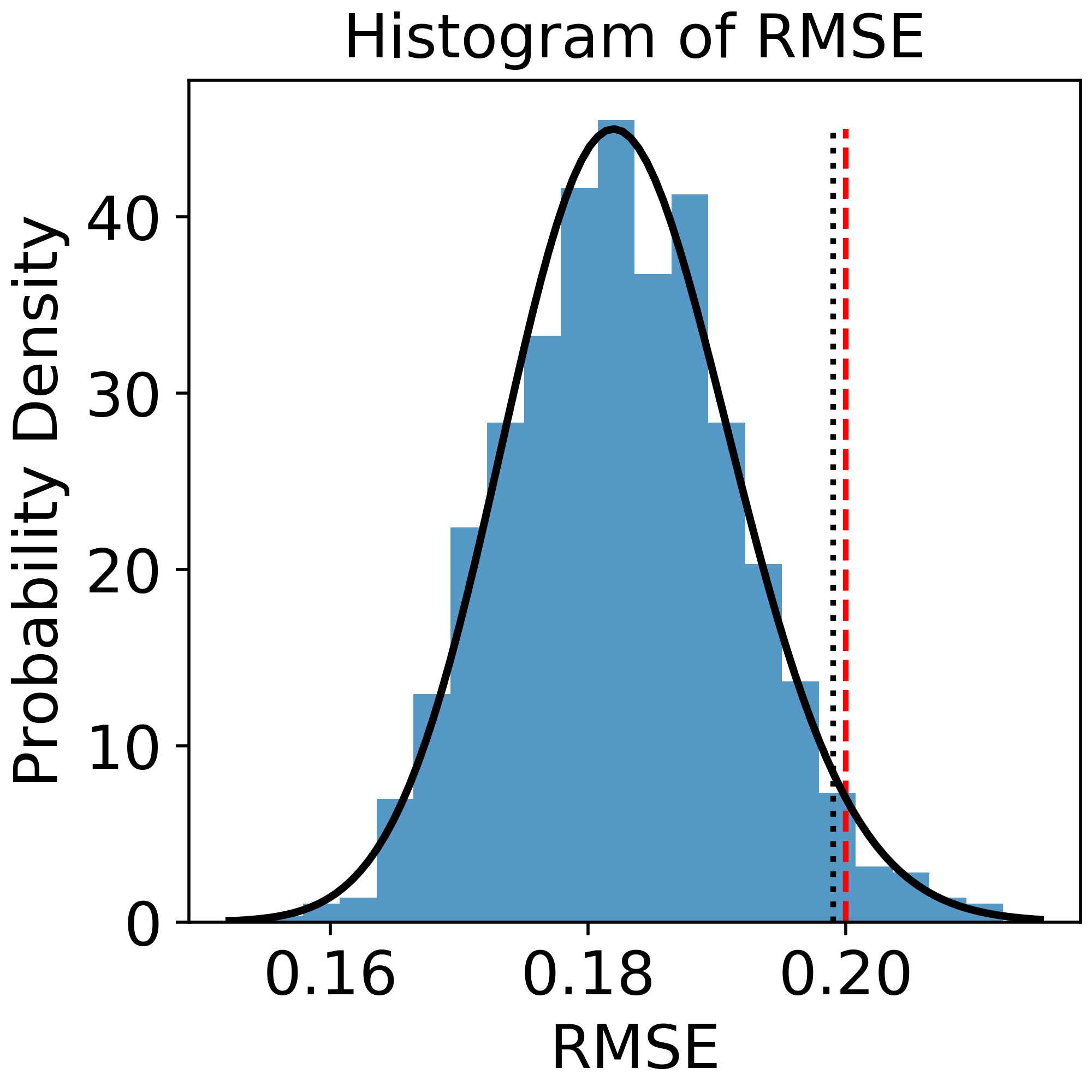}
    
    \caption{\footnotesize RMSE metric distribution for the run of our method that provided an inconclusive result once the experimental budget was exhausted. The grey dotted line shows the true RMSE value for this surface and the red dashed line is the threshold \(T_{avg}\).}
    \label{fig:RMSEIncon}
    \vspace{-50pt}
    \end{center}
\end{wrapfigure}

Interestingly, for \(n_{min} > 8\) and higher values of \(p_{accept}\), the algorithm failed to determine whether 1 network out of the 120 tested should be accepted or rejected within the experimental budget. The network in question has a true RMSE value very close to the threshold, so the method never became certain enough that the network should be accepted. Figure~\ref{fig:RMSEIncon} shows a plot of the RMSE metric distribution for this network with 50 observed data points.

The number of runs to termination and the nature of the termination for \(n_{min}=8\) are shown in the histograms in Figure~\ref{fig:experiments_term}. These results show that the stopping criterion can identify valid and invalid network architectures with few data points, and that the time to termination is dependent on the confidence in prediction required.

\section{Conclusion}

We have presented a method for determining if a network should be accepted or rejected for a given application using a sequential experimental approach and a stopping metric designed to give a measure of discrepancy across the entire surface. We tested this method on simulated biomolecular networks and demonstrated that it can accept or reject networks with a reasonable degree of accuracy in a small number of experimental runs.

Two main improvements will be made in future work. Firstly, we can use acquisition functions tailored for Bayesian quadrature, since we are trying to maximize our certainty about an integral. Recent work on modeling non-negative functions is particularly relevant \cite{gunter_sampling_2014, osborne_active_2012}. We also aim to eliminate the need for the minimum number of experiments parameter by integrating over the Gaussian process hyperparameters. 

\pagebreak

\begin{ack}

This work was supported by the UKRI CDT in AI for Healthcare \url{http://ai4health.io} [Grant No. EP/S023283/1], UK Research and Innovation [Grant No. EP/P016871/1] and US NIH [Grant No. 5F32GM131594]. 
\end{ack}

\bibliography{ML4MolesPaper}


\clearpage

\appendix

\section{Design Criteria Comparison}\label{ap:DesignCriteria}

\begin{figure}[h]
    \centering
    
    \begin{minipage}[b]{0.5\textwidth}
    \centering
     \includegraphics[scale=0.52]{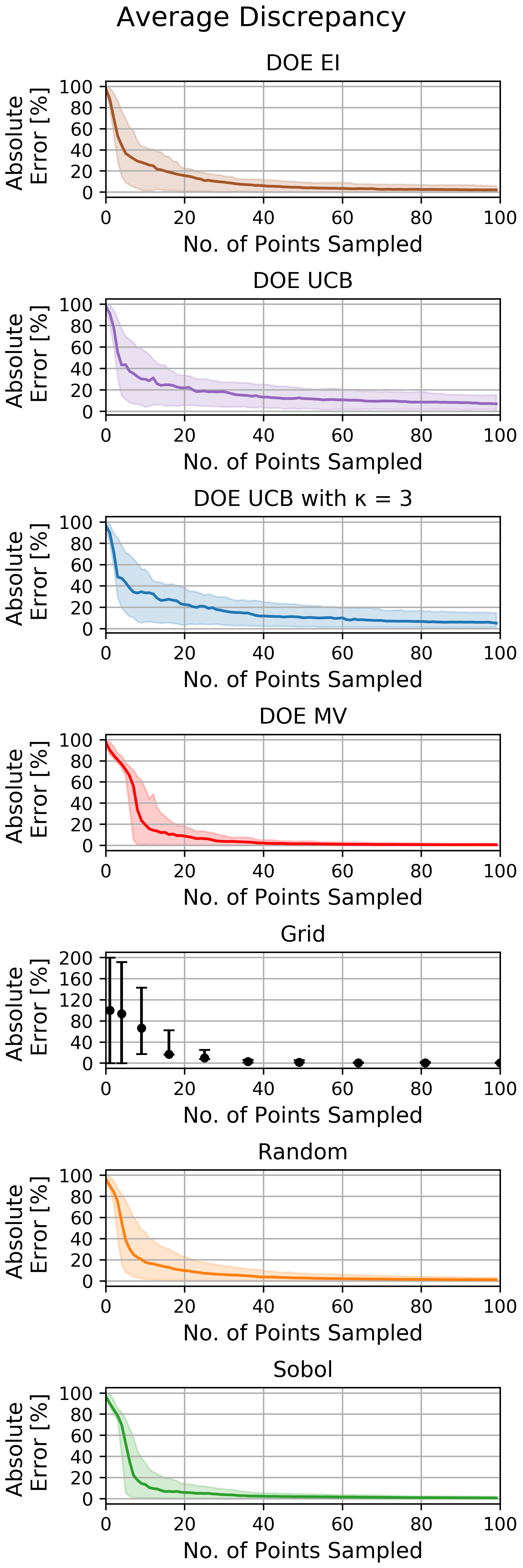}
    \end{minipage}
    \begin{minipage}[b]{0.49\textwidth}
    \centering
    \includegraphics[scale=0.52]{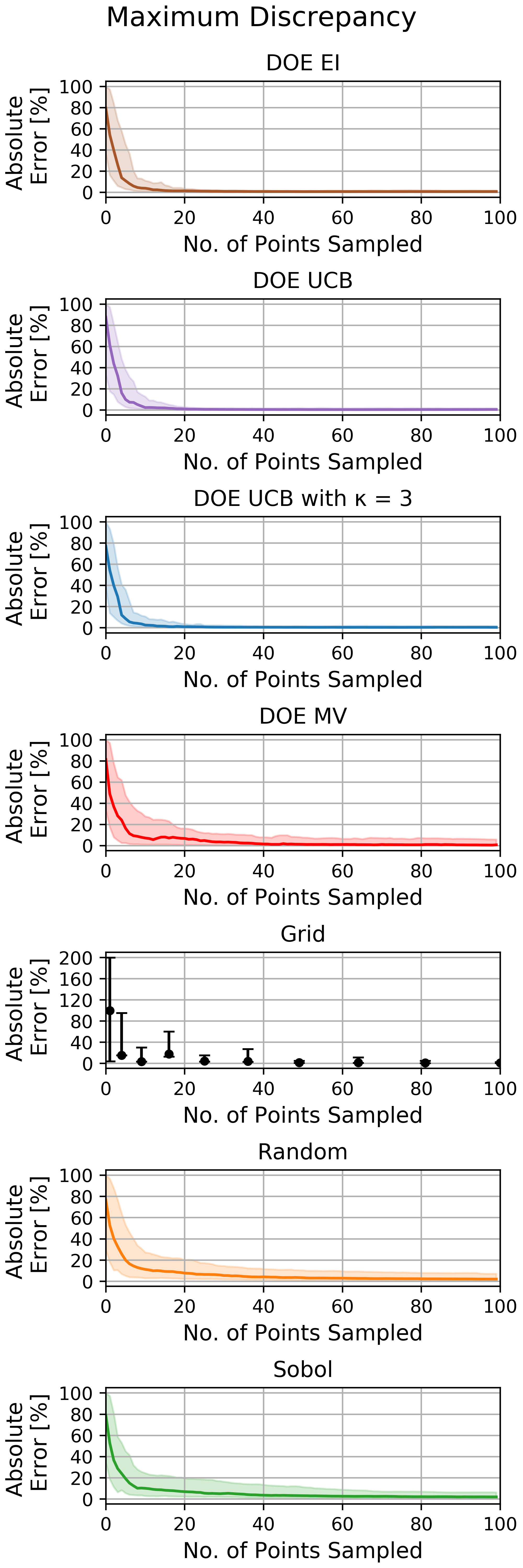}
    \end{minipage}
  
    \caption{\footnotesize The absolute error of the predicted mean and maximum discrepancy for 8 test functions compared to the true mean discrepancies for Sobol random, random, grid and the following design criteria: maximum variance, upper confidence bound (UCB), upper confidence bound with a higher weighting on the standard deviation term (UCB with \(\kappa = 3\)), and expected improvement (EI).  The grid search, random and Sobol methods were all run on each test surface 100 times. Due to time constraints, the DOE strategy with different design criteria were only run 20 times for each test function. For the grid method, the tests were only run for numbers of points that are square numbers, and it is assumed that for a given experimental budget, the number of points selected would be the largest square number within that budget. The solid line in each plot is the median value and the shaded region denotes the interquartile range.}
    \label{fig:design_crit_test_all_avg}
\end{figure}

\clearpage
\section{Biomolecular Networks Validation}\label{ap:biochem}

\begin{figure}[h]
    \centering

     \includegraphics[scale=0.4]{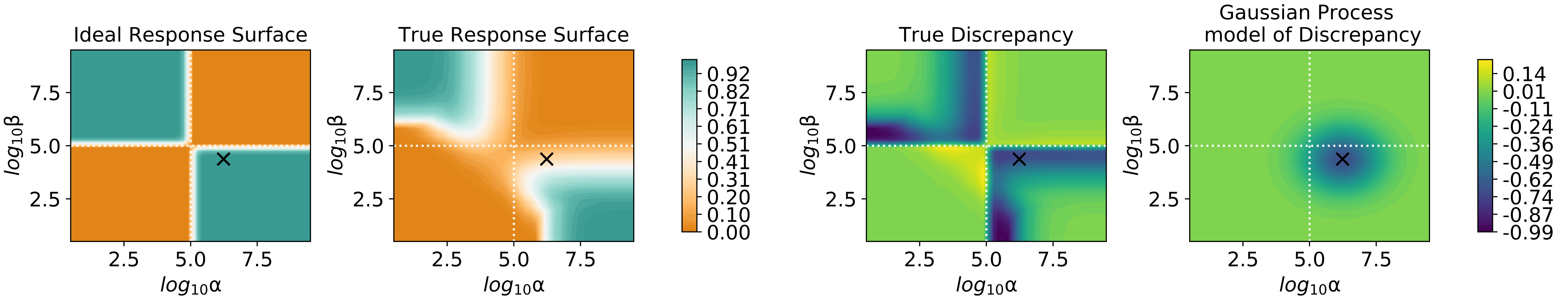}

     \begin{minipage}[b]{0.23\textwidth}
     \includegraphics[scale=0.41]{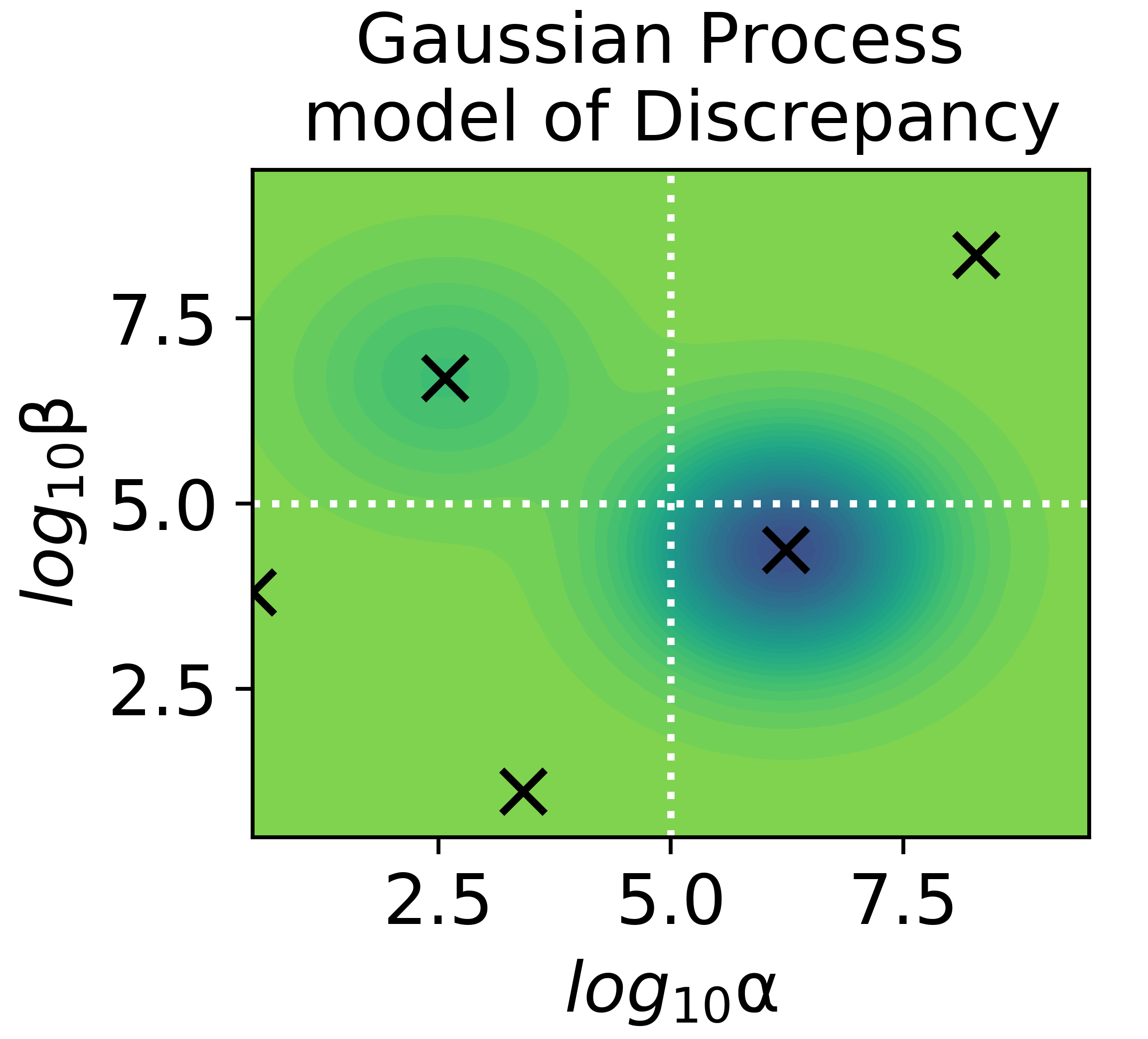}
     \end{minipage}
          \begin{minipage}[b]{0.23\textwidth}
     \includegraphics[scale=0.41]{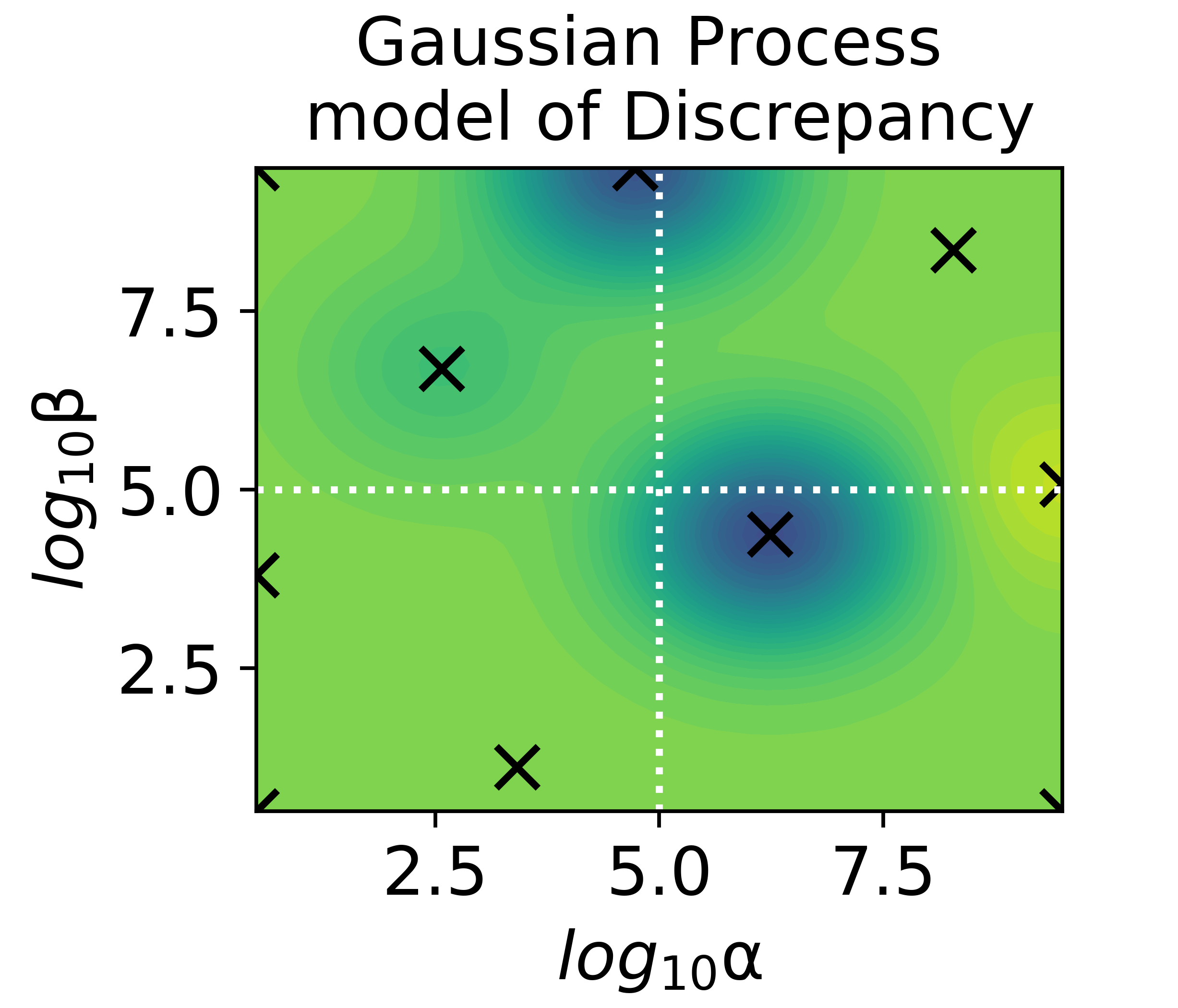}
     \end{minipage}
     \begin{minipage}[b]{0.23\textwidth}
     \includegraphics[scale=0.41]{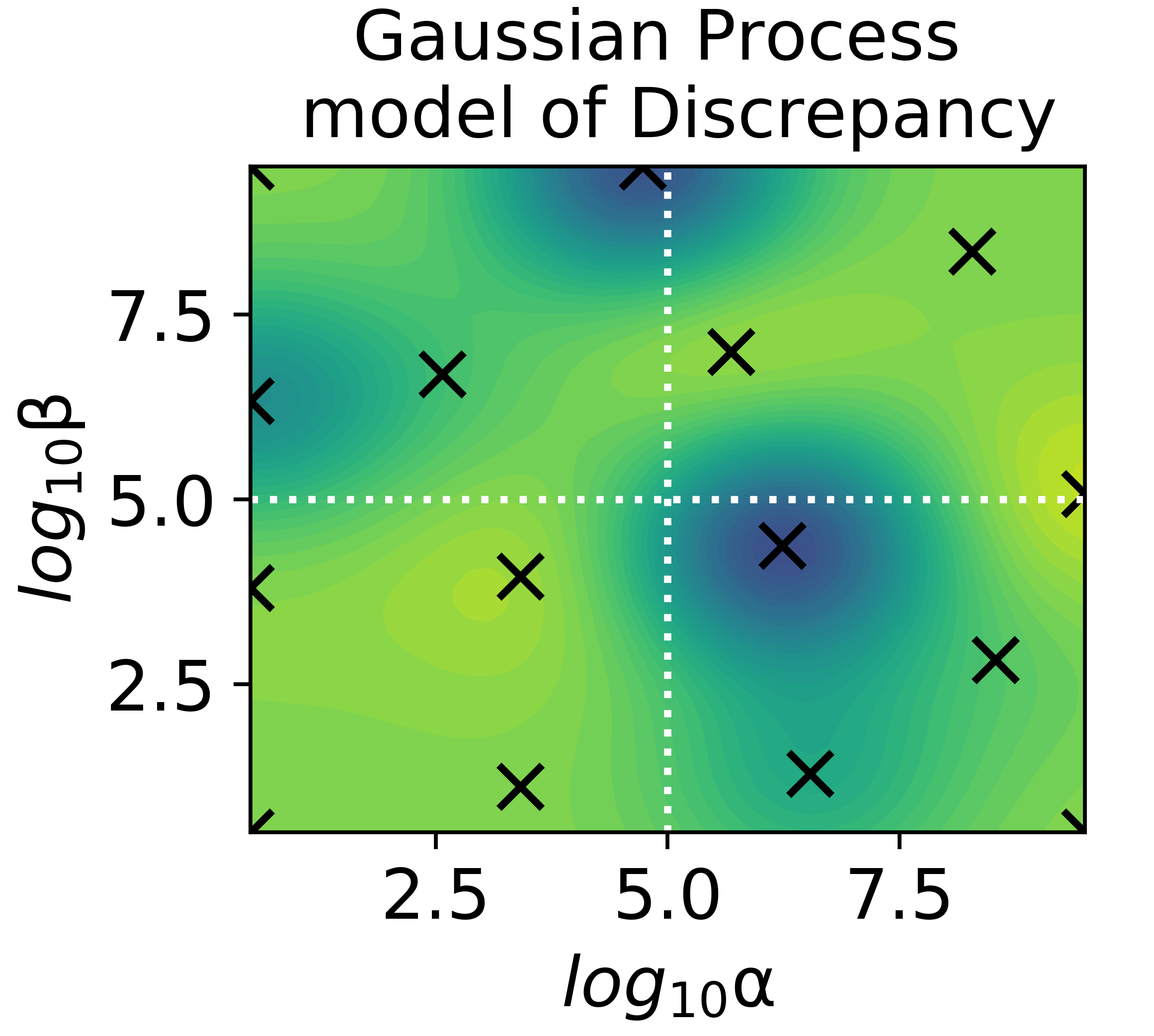}
     \end{minipage}
          \begin{minipage}[b]{0.29\textwidth}
          \centering
     \includegraphics[scale=0.41]{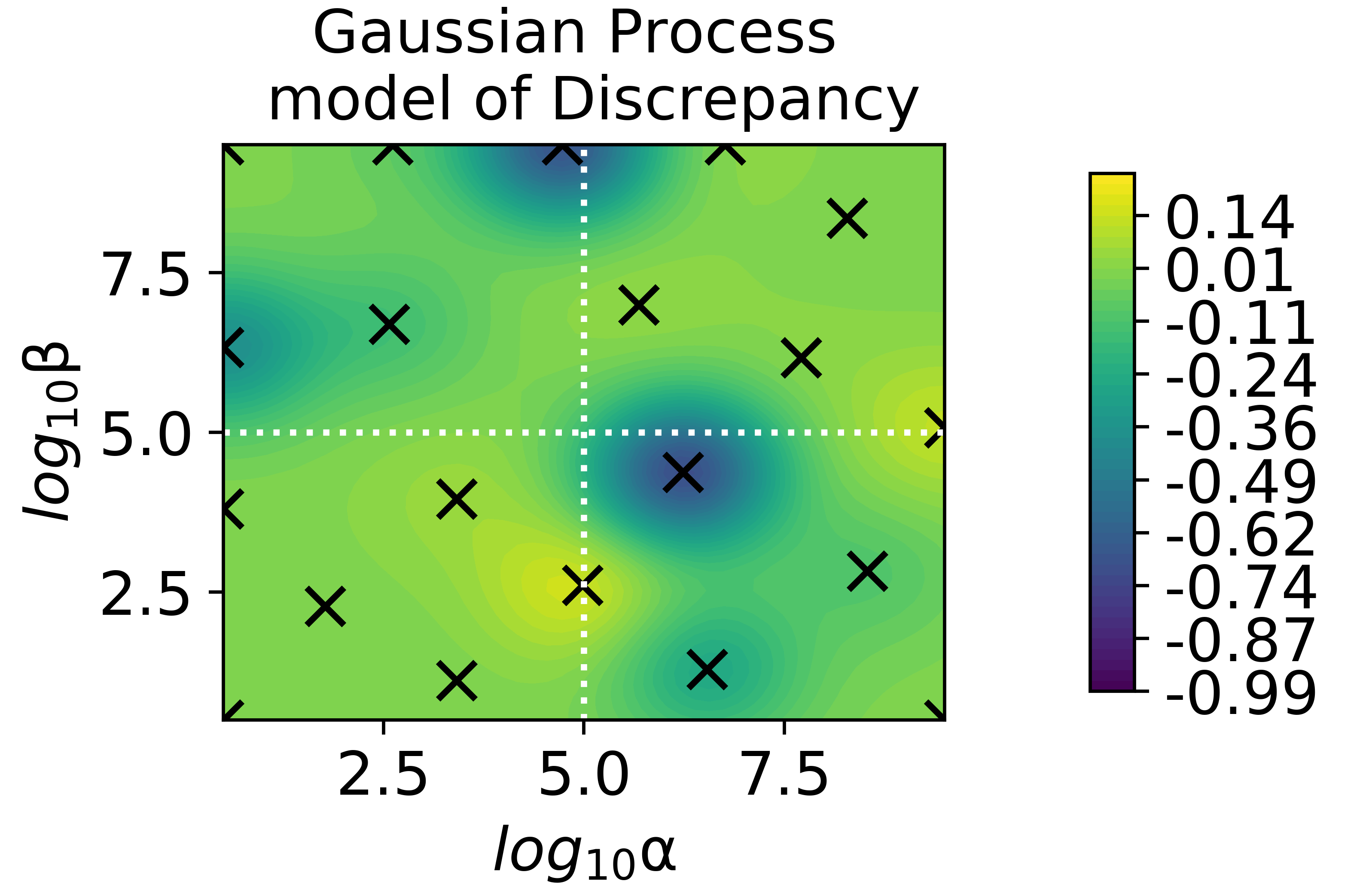}
     \end{minipage}

    \caption{\footnotesize An illustration of the sequential nature of the method. \(\alpha\) and \(\beta\) are input concentrations of biomolecules and the two plots in the top left hand corner show the ideal XOR response surface and the true response surface as generated by the biomolecular simulations. The plot second from top left shows the true discrepancy between these surfaces. The remaining plots then show the Gaussian process belief over the surface after 1, 5, 10, 15 and 20 experiments. The black crosses show where on the surface experiments have been conducted. Each new experimental point is selected using the maximum variance acquisition function.}
    \label{fig:sequential}
\end{figure}

\end{document}